\documentclass[manuscript, sigconf]{acmart}
\usepackage{enumerate}

\AtBeginDocument{%
  }

\usepackage{graphicx} 
\usepackage{xcolor}
\usepackage{tabularx}

\usepackage{vcell}
\usepackage[figuresright]{rotating}

\newcolumntype{L}{>{\RaggedRight\hspace{1pt}}l} 

\usepackage [autostyle, english = american]{csquotes}
\MakeOuterQuote{"}

\usepackage{hyperref}

\usepackage{setspace}

\setcopyright{acmlicensed}
\copyrightyear{2024}
\acmYear{2024}
\acmDOI{XXXXXXX.XXXXXXX}

\acmConference[Conference acronym 'XX]{xx}{June 03--05,
  2018}{Woodstock, NY}
\acmISBN{978-1-4503-XXXX-X/18/06}

\newcommand{\anedit}[1]{{\color{black}#1}}

\begin{document}

\title[Demystifying Technology for Policymaking]{Demystifying Technology for Policymaking: Exploring the Rideshare Context and Data Initiative Opportunities to Advance Tech Policymaking Efforts} 



\author{Angie Zhang}
\affiliation{%
  \institution{School of Information\\University of Texas at Austin}
  \city{Austin, TX}
  \country{USA}
  }

\begin{abstract}
In the face of rapidly advancing technologies, evidence of harms they can exacerbate, and insufficient policy to ensure accountability from tech companies, \textbf{\textit{what are HCI opportunities for advancing policymaking of technology?}} In this paper, we explore challenges and opportunities for tech policymaking through a case study of app-based rideshare driving. We begin with background on rideshare platforms and how they operate. Next, we review literature on algorithmic management about how rideshare drivers \textit{actually} experience platform features---often to the detriment of their well-being---and ways they respond. In light of this, researchers and advocates have called for increased worker protections, thus we turn to rideshare policy and regulation efforts in the U.S. Here, we differentiate the political strategies of platforms with those of drivers to illustrate the conflicting narratives policymakers face when trying to oversee gig work platforms. We reflect that past methods surfacing drivers' experiences may be insufficient for policymaker needs when developing oversight. To address this gap and our original inquiry---\textit{what are HCI opportunities for advancing tech policymaking}---we briefly explore two paths forward for holding tech companies accountable in the rideshare context: (1) data transparency initiatives to enable collective auditing by workers and (2) legal frameworks for holding platforms accountable.
\end{abstract}

\ccsdesc[500]{Human-centered computing~Human computer interaction (HCI)}
\ccsdesc[500]{Human-centered computing~Computer supported cooperative work}
\keywords{Gig Work, Rideshare Platforms, Tech Policymaking, Collective Auditing}


\maketitle

\section{Introduction}
AI and ML products are increasingly integrated into the everyday lives of the public. For example, automated decision-making tools are used by governments to make determinations around housing \cite{kuo2023understanding}, social services \cite{brown2019toward, saxena2020human}, and school assignment \cite{robertson2021modeling}. Personalized recommendation algorithms are employed by companies to curate user content \cite{gomez2015netflix}. And workplace technologies can screen candidates \cite{amad2015advantages} or manage worker productivity \cite{lomborg2022everyday}. These tools have gained popularity for their potential to improve efficiencies at large scale \cite{cunha2020deep}, reduce human bias in processes such as applicant selection \cite{chamorro2019will, lin2021engineering}, and automate time-consuming and redundant tasks \cite{dam2019towards}. However, they have also revealed the potential for problematic outcomes: algorithms intended for optimizing social service allocation can disparately impact different populations \cite{Ho_Burke_2023}; hiring algorithms have been found to reinforce racial and gender biases \cite{engler2021auditing, dastin2022amazon}; companies' collections of user data for personalized recommendations have enabled predatory online marketing schemes \cite{jan2019hud}; and generative AI is intensifying security risks such as realistic AI-enabled voice scams \cite{Sabin_2023}. 

Given this evidence of how unregulated technology can and has impacted people negatively, calls are increasingly being made to demand regulation and policy to hold companies and their technologies accountable. The need for comprehensive tech regulation and policy is of particular importance for upholding labor rights. As technology-mediated work proliferates and evolves, policies specifying requirements around algorithm disclosure, data transparency, and privacy continue to lag. Meanwhile, the unchecked algorithmic management of app-based gig work platforms exposes workers to opaque, harmful practices including: undisclosed A/B testing that can reduce a worker's earnings \cite{deb2018under, muller2019algorithmic}; unclear work assignment or termination appeal methods that can lead to lost wages while workers' cases are researched without their input \cite{lee2015working, myhill2021job}; \anedit{algorithmic pricing} and variable platform commissions decreasing worker wages \cite{konishi2023analysis, myhill2021job}; and confusing contract language deceiving workers into ceding rights to arbitration \cite{Lien_2015}.

Although researchers, workers, and advocates have sought to make evident the ways algorithmic management impacts workers---e.g., the manipulation workers experience, the behavioral responses they adopt in order to overcome it, and the negative effects algorithmic management can have on worker well-being---for various factors, policymakers and government agencies may still face challenges in understanding and linking worker harm to the algorithmic management that causes it, and \anedit{what actions---e.g., legislative, regulatory, legal---}are necessary. First, in response to worker grassroots organizing for better pay and treatment, platform lobbyists wield political clout to muddy organizers' demands by discounting workers' experiences and threatening consumer price hikes \anedit{or market exit} to intimidate policymakers and regulators \cite{stein2020analysis, Cummings_2023}. Additionally, platforms often create specialized terminology for the algorithms workers are subjected to, perplexing policymakers and complicating efforts of researchers or advocates attempting to explain these concepts and their impacts on workers. Furthermore, policymakers often seek evidence---i.e., data---to support the need for policy \cite{manuel2017participatory}. Yet, the data required for analyzing platform practices and the effects on workers has been difficult to obtain due to companies claiming trade secrets \cite{gutman2018uber, mohlmann2019people}. All of this compounds and complicates for policymakers whether and how to craft policy for holding platforms accountable. 

Often, bureaucratic roles such as policymakers, regulators, and government agencies wield exclusive formalized power to hold companies responsible for fair worker conditions and transparent work algorithms. Thus, it is crucial that these stakeholders can clearly understand how platforms operate and connect it to the impacts on workers. To that end, we ask:

\begin{quote}
    \textbf{What HCI opportunities exist to advance policymaking of technology?}
\end{quote}

This is pertinent given recent calls by researchers for the HCI and CSCW communities to play a more prominent role in shaping not only the design of technology, but simultaneously, the design of technology policy \cite{manuel2020place, yang2023designing, spaa2019understanding, clemmensen2021hwid, jackson2014policy, jackson2013cscw, lazar2016human}. 

\subsection{Roadmap}
In order to understand how to support technology policymaking for 
workers, we focus on a specific form of app-based gig work: rideshare driving. Despite existing for more than a decade, regulations and policies in the United States have largely been nonexistent at the federal level and inconsistent at the state and municipal levels. Thus, we explore the literature to answer the following questions: 

\begin{enumerate}
    \item \textbf{What challenges exist for policymakers and regulators when overseeing rideshare platforms?} 
    \item \textbf{What HCI opportunities exist for advancing policy and regulation related to app-based rideshare companies?}
\end{enumerate}

To answer the first question, we explain how rideshare platforms function to highlight their motivations and business strategies (Section \ref{Section_Background_Rideshare}). Next, we review algorithmic management literature to contrast how app features are presented by platforms with how they actually manipulate and control workers (Section \ref{Section_algorithmic_management}). In light of this, advocates and researchers have called for worker protections, thus we turn to rideshare policy and regulation efforts in the U.S.: here, we juxtapose the political strategies of platforms with those of drivers to convey the complexities policymakers face when trying to oversee gig work platforms (Section \ref{Section_policymaking}).

We reflect that while research has surfaced the experiences of app-based gig workers, this work is often conducted in a way that makes it difficult to directly connect worker anecdotes or survey results with \textit{how platforms engendered these}. 

\anedit{To answer the second question, we briefly describe two potential directions for HCI and policy/regulation collaborations (Section \ref{Future}). First, data transparency has increasingly been of interest to researchers for its promise to enable auditing that holds technology companies accountable for its worker and consumer practices. We briefly outline tools and efforts aligned with empowering workers to raise issues with local legislators via data advocacy and auditing. Second, important work is being done by scholars to develop policy or legal foundations for holding platforms accountable, such as \citet{dubal2023algorithmic}'s work on algorithmic pricing and wage discrimination. We consider this and discuss ideas of how HCI contributions may support this and other frameworks. Finally, we share resources we have compiled\footnote{Note: These are not intended to be comprehensive resources, as new reports, terminology, policy/regulation events may arise. But we provide them in hopes they may be useful for others interested in exploring these topics.} including:
}
\begin{enumerate}
    \item A friendly, companion website to this paper for anyone to read, where users can also explore a Work Planner data probe (e.g., try our calculator to view how much you might make in a week as a rideshare driver!)

    \href{https://www.demystifying-gigwork.com}{\color{blue}Website: demystifying-gigwork.com}.
    
     \item A simple compilation of related rideshare policy/regulation reports by worker-organizer groups, economists, governments, and platforms.
    
    \href{https://airtable.com/app09qGH7CDmoN6qE/shrQEYxZjSUYQrZy7/tblUBrABNUUToPVfN}{\color{blue}Link to Repository of Reports}.
    
    \item A timeline of rideshare (and other relevant app-based gig work) policy/regulation related events.
    
    \href{https://airtable.com/appZvP15Nt6EAjV8R/shrBZ9AKDhJ7woa9s/tblVguO09iLEE18o8}{\color{blue}Link to Timeline}.

    \item A glossary of related (rideshare gig work) terms.
   
   \href{https://airtable.com/appMlzch3KfhPzK2R/shrKrHSstyAIS3usV/tblQ1XsbYafev121J}{\color{blue}Link to Glossary}.
   
\end{enumerate}



\section{Background on Rideshare Platform Operations: You Can't Regulate What You Don't Understand}\label{Section_Background_Rideshare}
It is necessary to understand how tech companies operate in order to create effective policy or regulation. According to workers and advocates, policymakers may not always take the time to truly understand how platforms work \cite{zhang2024data}. Thus, we provide background information about how rideshare platforms operate in an effort to shed light on the status quo they wish to maintain.  

\subsection{Introduction of Rideshare Platforms}
One of the most popular forms of on-demand app-based gig work is rideshare driving, largely dominated by two companies in the United States---Uber and Lyft---since 2009 and 2012, respectively. When it launched, Uber (then \textit{UberCab}) was hailed for its use of big data to bring together riders and drivers through its powerful matching algorithm \cite{Chen_2011}. The two platforms have grown rapidly since: Lyft operating in 600+ cities across the US and Canada with 1.4 million drivers\footnote{https://therideshareguy.com/lyft-statistics/}, and Uber in 10,000+ cities across 72 countries with over 5 million drivers\footnote{https://therideshareguy.com/uber-statistics/}. Although the companies tout multitudes of workers, they maintain that they are not job creators or employers, and rideshare driving remains a largely unregulated industry in the United States, despite its surface similarity to the highly regulated taxicab industry \cite{dubal2017drive}. 

Through the rideshare platform, individuals can sign up to be drivers and/or customers. Approved drivers use a separate driver-side app where the platform will send them ride requests with select trip information, and the driver must accept or reject the trip, often within a few seconds. Customers can use the passenger-side app to input desired pick-up and drop-off locations, view fare prices, request a ride, and be matched with a driver by the platform. Fares are set by platforms, not drivers. Platforms make money by charging a commission on the fares that customers pay to the drivers they have been matched to, and customers are not told how much of their fare goes to drivers nor are drivers told how much the customer paid in total. To maximize their profit, platforms seek to increase their number of passenger bookings by providing customers with short wait times and reasonable fares. In order to ensure low wait times and fares, rideshare platforms require a large fleet of available (i.e., idle) drivers to balance the supply and demand of trips and employ various methods to compel drivers to come "online". 

\subsection{Platform Operations \& Terminology} \label{subsection_terms}
Platforms often create specialized terms to describe application features they use to manage workers. 
We describe a subset of these terms to aid the reader's understanding about how platforms operate and make money. We will revisit these terms and expand on how \textit{workers} experience platform tactics in Section \ref{subsection_tactics}.


\textbf{Bonuses: Quests (Uber)\footnote{https://www.uber.com/blog/how-quest-works}, Challenges (Lyft)\footnote{https://help.lyft.com/hc/en-us/all/articles/360001943867-Ride-Challenges}}. \textbf{Quests} are Uber's volume-based promotions where drivers can earn a bonus in addition to what they earn after each ride. A driver must complete a preset number of trips in a specific time range, e.g., Monday through Friday 5am-5am. The platform determines all parameters of the Quest, including how many trips must be completed, the bonus amount awarded, the time span the trips must be completed within. Quest offers can differ between drivers and a driver is not guaranteed to receive the same Quest offer again. The method for determining what Quest a driver will receive is not disclosed. Typically, Quest bonuses are not prorated: if a driver does not complete the minimum amount of rides for a Quest, they will not receive any of the bonus. Quests are advertised by Uber as a way for drivers to maximize their earnings and allow the company to increase driver supply on the roads, decrease passenger wait time, and balance fares. Lyft's equivalent is \textbf{Challenges} and operates similarly to Quests. 

\textbf{Driver Loyalty Tiers: Uber Pro (Uber)\footnote{https://www.uber.com/nz/en/drive/uber-pro}, Lyft Rewards (Lyft)\footnote{https://www.lyft.com/driver/rewards}}. Uber Pro is Uber's loyalty program intended to reward high-performing drivers. Drivers are automatically enrolled in the base tier, "Blue", and can ascend to the next three (Gold, Platinum, Diamond) by earning points. Points are obtained from driving and are aggregated over 3-month periods to determine the driver's current status. The higher their tier, the more access drivers have to off-app perks such as discounts on driving-related services and on-app perks such as a higher degree of trip information disclosure.  A driver can be penalized and lose their tier if they 1) cancel too many trips, or 2) receive too many sub-5 star reviews. Lyft's equivalent is called \textbf{Lyft Rewards}. 

\textbf{Dynamic Pricing: Surge Pricing (Uber)\footnote{https://www.uber.com/us/en/drive/driver-app/how-surge-works/}, Prime Time (Lyft)}\footnote{https://ride.guru/content/newsroom/how-to-navigate-lyfts-prime-time-fares}. In order to balance driver supply and customer demand, Uber uses dynamic pricing, "surge pricing", to charge passengers a premium and increase fares for drivers to incentivize them to work. Surge pricing can be specific to a time of day or a geographic region in which demand is particularly high (e.g., hours of workday starts and ends, special events, inclement weather). For example, if surge pricing is specific to a geographic area, platforms will display heat maps to drivers on the driver-side app of where surge pricing is occurring. If a driver is in an active surge pricing zone or receives a trip request that picks up in a surge area, a multiplier will be applied to the fare's base rate, thereby increasing the driver's earnings. Lyft's equivalent is often referenced to as surges, but they also use the term \textbf{Prime Time} to refer to times when drivers are eligible to receive incentives on top of their base pay. 

\textbf{Fare Calculation: Upfront Pricing (Uber)\footnote{https://www.uber.com/us/en/marketplace/pricing/upfront-pricing/}, Upfront Pay (Lyft)}\footnote{https://www.lyft.com/blog/posts/upfront-pay-and-the-next-chapter-of-the-lyft-driving-experience}. When Uber and Lyft first debuted, fares were calculated on a fixed per mile and per minute rate for the time and distance driven between a passenger's pick-up location to drop-off destination, and platform commissions were a constant 25\%. In the last few years, both platforms have switched to an algorithmically variable pricing model. Platforms explain this change has increased transparency for drivers because they can now view information about trip requests not previously disclosed (e.g., passenger pick-up and drop-off locations in some cities)\footnote{\label{upfrontuber}https://www.uber.com/us/en/drive/how-much-drivers-make/?uclick\_id=483be887-4d6c-45c2-bfbb-ae42ddab018f}. Platforms also state that this pricing model can potentially incorporate payment for work time previously unaccounted for, such as driving to pick up a passenger\footnotemark[\getrefnumber{upfrontuber}]. These new pricing models do not disclose details to drivers or passengers about how fares are calculated.



\section{Impacts of Gig Work Platforms on Workers} \label{Section_algorithmic_management}
Platforms often frame rideshare driving as an attractive option for workers due to low barrier of entry, the opportunity to "be your own boss", and the allure of flexibility to "work when you want". A closer look reveals "algorithmic management" supersedes worker autonomy. This term, coined by \citet{lee2015working}, explains how in the absence of human managers, platform algorithms assign work, calculate wages, and determine performance and termination \cite{lee2015working}, as well as influence temporal and spatial movement \cite{shapiro2020dynamic}. Additionally, social science and HCI research have repeatedly surfaced that platforms exert power and information asymmetries, such as concealing how variable pricing is determined or withholding essential trip information from drivers before they accept them \cite{rosenblat2016algorithmic, gloss2016designing, lee2015working, wood2021platform} at the expense of drivers' physical, financial, and psychological wellbeing \cite{gloss2016designing, wood2021platform, zhang2022algorithmic, hill2021algorithmic, attoh2019we, mohlmann2021algorithmic}. 

\subsection{Characterizing Driver Behaviors}
In response, drivers employ strategies of varying defiance---from resistance \cite{mohlmann2017hands} or deviance \cite{cameron2020rise} such as liberally rejecting or cancelling underpaying or unsafe trips; gaming \cite{mohlmann2017hands} or engagement \cite{cameron2020rise}, where drivers may collectively make sense of the system to manipulate it; switching \cite{mohlmann2017hands} or working multiple apps simultaneously to maximize up-time; and compliance \cite{cameron2020rise} or allowing platform features to shape their decision-making. \citet{dubal2023algorithmic} characterizes these drivers' behaviors akin to a gambling mindset, troubling as it results in drivers trying to "win" at a system designed to favor platforms as evidenced by their ever-changing nature \cite{Kerr_2022, paul2020uber, mohlmannn2023algorithm}, often to drivers' financial detriment \cite{dubal2023algorithmic, Sherman_2023}. Earlier in Section \ref{subsection_terms}, we described the mechanics of how a subset of platform features function. Now, we demonstrate how these same features are actually used by platforms to facilitate algorithmic management by describing how workers experience them.

\begin{table*}[]
\begin{tabular}{|c|c|c|l|l|}
\hline
\multicolumn{1}{|l|}{\textbf{Feature}}                            & \multicolumn{1}{l|}{\textbf{Uber}}                        & \multicolumn{1}{l|}{\textbf{Lyft}}                             & \textbf{Platform Characteristics}                                                                                                                                                                                                                                                                                                                                                                                         & \textbf{Worker Experiences}                                                                                                                                                                                                                                                                                                                                                                       \\ \hline
\begin{tabular}[c]{@{}c@{}}Bonuses\\ or\\ Promotions\end{tabular} & Quest                                                     & \begin{tabular}[c]{@{}c@{}}Challenges, \\ Streaks\end{tabular} & \begin{tabular}[c]{@{}l@{}}Allows drivers to earn a bonus \\ on top of their normal fare earnings.\\ \\ Drivers must complete a \\ specific \# of trips within a \\ specific time range. \\ Platforms decide the \# of trips, \\ time period, and bonus amount.\end{tabular}                                                                                                                                               & \begin{tabular}[c]{@{}l@{}}Quests are unfairly distributed across drivers, \\ and designed to be unattainable \cite{zhang2022algorithmic}.\\ \\ Quests require long hours to complete, \\ and after completion, subsequent Quest\\ offers require drivers to work longer for \\ less pay \cite{vasudevan2022gamification, zhang2023stakeholder, mason2018high}. \end{tabular}                                                                                                                        \\ \hline
\begin{tabular}[c]{@{}c@{}}Loyalty\\ Tiers\end{tabular}           & \begin{tabular}[c]{@{}c@{}}Uber\\ Pro\end{tabular}        & \begin{tabular}[c]{@{}c@{}}Lyft\\ Rewards\end{tabular}         & \begin{tabular}[c]{@{}l@{}}Rewards drivers who meet platform-\\ defined criteria with benefits in-app \\ or perks outside of the app \\ (e.g., viewing more detailed trip \\ information, receiving discounts on fuel).\\ \\ Drivers must drive a minimum \# of \\ trips and maintain a driver rating \\ and cancellation rate above a certain \\ threshold.\end{tabular}                                                  & \begin{tabular}[c]{@{}l@{}}Drivers are penalized for exercising more \\ autonomy when they are more selective \\ with trip acceptances \cite{page2017perceived, borowiak2021algorithm, sehrawat2021everyday}.\\ \\ Compels drivers to accept most or all rides---\\ potentially non-preferential or dangerous trips---\\ in order to maintain high acceptance rates\\ \cite{zhang2022algorithmic, vasudevan2022gamification}.\end{tabular}                                                                                       \\ \hline
\begin{tabular}[c]{@{}c@{}}Dynamic\\ Pricing\end{tabular}         & \begin{tabular}[c]{@{}c@{}}Surge\\ Pricing\end{tabular}   & \begin{tabular}[c]{@{}c@{}}Prime\\ Time\end{tabular}           & \begin{tabular}[c]{@{}l@{}}Pays drivers a premium for driving \\ during peak times or in neighborhoods \\ of particularly high demand.\\ \\ Drivers must be working at the \\ precise time and/or location that a \\ surge pricing is applied to. Surge \\ pricing is updated in real time and \\ can change quickly.\end{tabular}                                                                                        & \begin{tabular}[c]{@{}l@{}}Drivers "surge chase" unsuccessfully due to \\inconsistent and unreliable information \\provided by apps \\\cite{lee2015working, rosenblat2016algorithmic, cameron2020rise, dubal2023algorithmic, mohlmann2017hands, ashkrof2020understanding}.\\ \\ Apps use notifications to pressure or nudge \\drivers to begin or continue working \\ without the guarantee of rides or earnings \\\cite{rosenblat2016algorithmic, cameron2020rise, cameron2022making, attoh2019we} .\end{tabular}                                                                                                    \\ \hline
\begin{tabular}[c]{@{}c@{}}Fare\\ Calculation\end{tabular}        & \begin{tabular}[c]{@{}c@{}}Upfront\\ Pricing\end{tabular} & \begin{tabular}[c]{@{}c@{}}Upfront\\ Pay\end{tabular}          & \begin{tabular}[c]{@{}l@{}}Provides drivers more information per \\ trip request before acceptance, \\ including location of passenger pick-up\\ and drop-off.\\ \\ Drivers can view more details when they \\ receive a ride request and are told their \\ effort spent picking up a passenger is \\ also included. Specific calculation details \\ of the fare are not provide to drivers \\ or passengers.\end{tabular} & \begin{tabular}[c]{@{}l@{}}Drivers describe increased opacity as well as \\variable and often high platform commissions \\or "take rates" than the previously fixed \\25\% platform commission \\\cite{dubal2023algorithmic, zhang2024data, Kerr_2022}.\\ \\ Drivers suspect platforms are using personalized \\dynamic algorithms to determine the maximum \\ passengers are willing to pay and the \\minimum drivers will accept for a ride \cite{zhang2024data}.\end{tabular} \\ \hline
\end{tabular}
\caption{\label{tactics_new2}This table displays information from Section \ref{subsection_terms} and \ref{subsection_tactics} for a more direct comparison.}
\end{table*}

\subsection{Platform Tactics and Impacts on Workers}

\label{subsection_tactics}
Workers' experiences with platform features illustrate how platform management and manipulation occur due to platform-induced information and power asymmetries, and gamification. Much of this analysis has been completed through qualitative means with drivers---ethnographic studies, driver interviews, analysis of driver forum posts or social media posts, workshops, and co-design sessions---as well as observations of platform features (e.g., driver screenshots of the driver-app) and platform ads. For consistency, we continue to describe these using terminology of Uber's platform as in Section \ref{subsection_terms}; however, the overarching driver experiences on Uber are translatable to Lyft. This comparison is also summarized in Table \ref{tactics_new2}.


\textbf{Bonuses: Quests.} Although Quests are framed by platforms as opportunities for drivers to drive more and earn more, drivers often describe Quests in reality as resulting in them working more for less pay \cite{vasudevan2022gamification, zhang2023stakeholder}. Drivers also perceive Quests as unfair based on personal and collective observations that newer or inactive drivers receive more "lucrative" promotions than loyal, consistent drivers \cite{zhang2022algorithmic}; and manipulative for they suspect platforms use dynamic algorithms to deliberately prevent drivers from completing Quests as they approach the required number of rides \cite{zhang2022algorithmic}. Given that platforms collect huge swaths of worker data to personalize gamification tactics such as individualized Quest promotions across workers based on their work history \cite{scheiber2022uber, mohlmann2021algorithmic}, this suspicion that their work data may also be used to hinder their progress is not entirely unfounded. Analyses of discussion board posts echo these unfair and manipulative properties of Quests \cite{pregenzer2021algorithms, vasudevan2022gamification},  with \citet{vasudevan2022gamification} observing a tension between drivers: while some posted about their frustration in the effort required of Quests and disproportionate payout, others viewed this as a "hustle" as acceptable and fundamental characteristics of being self-employed workers. 


\textbf{Driver Loyalty Tiers: Uber Pro.} Loyalty tiers like Uber Pro are portrayed as programs to reward long-term drivers, yet drivers explain how tiers pressure and control their behavior instead. Drivers in \cite{zhang2022algorithmic} explained the contradictory nature of rewards: ascending to a higher tier allows drivers to view more information about ride requests, yet if they act on this new data to be more strategic, they are punished. For example, if a driver wants to maximize short trips and is now in a tier that allows them to view trip times, they could reject trips above a certain duration. However, turning down or cancelling too many rides can also lead to drivers being penalized or deactivated \cite{page2017perceived, borowiak2021algorithm, sehrawat2021everyday, attoh2019we}.

\citet{vasudevan2022gamification} observed how tiers reinforce the idea that basic information workers should have access to must be earned, echoing the comments of a driver in \cite{zhang2022algorithmic} who expressed "They [Uber] definitely treat [trip information] like it’s a perk when it’s kind of a necessity." This acts to steer driver behaviors away from what they may want to do---be more selective about what trips they accept and liberally reject requests---to accept most or all rides in order to maintain high acceptance rates \cite{maffie2023visible}, a metric platforms want drivers to maximize. 


\textbf{Dynamic Pricing: Surge Pricing.} Despite surge pricing presented by rideshare companies as opportunities for drivers to earn more, early research by \citet{lee2015working} and \citet{rosenblat2016algorithmic} already began to surface drivers' experiences with surges as faulty and unreliable. While some drivers of \cite{lee2015working} spoke about using surge pricing maps to direct where they wait for rides, more than half reported not altering their behavior due to unsuccessful experiences pursuing surges. Driver interviews and driver forum posts analyzed by \citet{rosenblat2016algorithmic} surfaced similar experiences: that drivers will enter surge zones of increased rates by "3.5x, but receive ride requests at a lower surge rate, such as 1.5x"; that they may wait in a surge zone only to receive ride requests in other, non-surge zones; that surges can disappear without warning. \citet{mohlmann2017hands} found even when drivers receive a trip with a surge bonus, the system may not automatically provide the increased fare and drivers have to independently pursue algorithmic and human-intervention support. 

Drivers have also described surge pricing as pressuring them to begin or continue working, with platforms sending them incessant notifications about active surges regardless of the drivers' preferences for using it to guide their strategies \cite{rosenblat2016algorithmic, cameron2022making, attoh2019we, scheiber2022uber}. \citet{rosenblat2016algorithmic} describe this as an exertion of "soft control" platforms use to nudge drivers to become available "without being guaranteed paid work", comparable to precarious scheduling often seen in shift work \cite{lambert2019precarious}.


\textbf{Fare Calculation: Upfront Pricing.} At the time of writing, Upfront Pricing is the most contemporary feature implemented on rideshare platforms of those discussed here. In 2022, Uber stopped calculating fares and platform commissions based on fixed rates, and started using an algorithm with undisclosed variables to make determinations. While platforms justified this move as increasing transparency because drivers can now view the estimated fare and the passenger's pick-up and drop-off locations prior to accepting a trip, some drivers have reported lower earnings and higher platform commissions (also called "take rates" or "platform cut" by drivers) \cite{Kerr_2022}. Driver organizers in \cite{zhang2024data} theorize this is happening because through Upfront Pricing, platforms can underhandedly and dynamically determine the maximum passengers are willing to pay and the minimum drivers are willing to work for based on their personal histories to increase platform take rates. Drivers interviewed in \cite{dubal2023algorithmic} described it as "a pay cut in disguise" with \citet{dubal2023algorithmic} arguing that Upfront Pricing and other pricing and incentive schemes (e.g., Quests, Surges) act as levers of wage control on workers, in order to support platform imposed "algorithmic wage discrimination"---that is, that platforms use personalized, dynamic algorithms to calculate disparate worker wages despite performing similar work.

\section{Rideshare Policymaking \& Regulation: A Clash Between Platforms and Driver Organizers}\label{Section_policymaking}
In the previous two sections, we explained how companies position themselves as software providers that match independent contractors with clients whereas rideshare drivers actually experience heavy-handed algorithmic management that controls their work, impacting their autonomy and well-being. Likewise, government oversight of rideshare companies can be characterized as an ongoing face-off between platforms attempting to convince policymakers to formalize their unregulated status as an industry and driver organizations demanding policymakers hold tech companies accountable for fair work practices. The overall goals of the two groups are usually at odds with one another, contributing to policymakers' confusion and frequently devolving labor rights into a political quandary.

\begin{figure*}
\includegraphics[width=.9\linewidth]{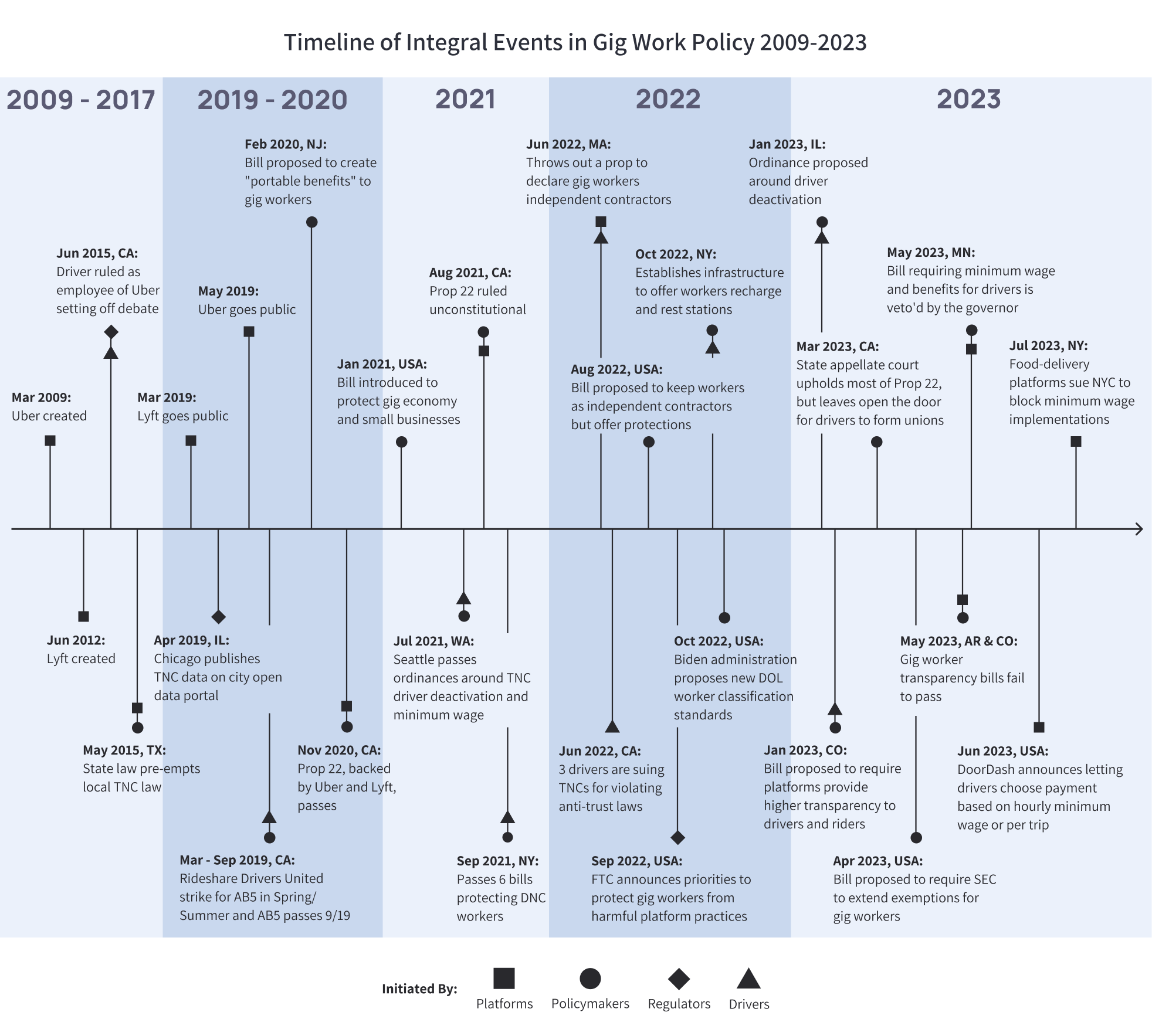}
  \caption{\textbf{An abbreviated overview of events in the United States related to app-based gig work policy and regulation evolution.} \normalfont Noticeably, there is a large gap between 2009-2017 of meaningful attempts, with more efforts occurring in 2022-2023. The endpoint(s) of each line denotes who the event was initiated by and/or benefited. For example, the passage of Prop 22 is denoted with a square (platform) and circle (policy) as it was a platform-backed policy that passed. In some cases, a policy does not have a corresponding square (platform) or triangle (worker groups) because it resulted in mixed outcomes for both groups.
  }
  \label{fig:timeline}
\end{figure*}

\subsection{Platform Objectives for Rideshare Regulation} 
The years immediately following their launch, Uber and Lyft were met by rare instances of pushback but otherwise relatively little resistance. However, in 2013, California became the first state to regulate rideshare driving by mandating universal background checks and a corporate insurance policy, among other changes, beginning a long journey of inconsistent and reactive regulation \cite{geron2013california}. Other local city governments attempted to follow suit, with early legislation mainly taking the form of background checks, fingerprinting, and minimum pay standard laws. However, platforms retaliated by threatening to leave markets---e.g., Seattle (2017), California (2020) \cite{Hawkins_2020}, and Minneapolis (2023) \cite{Henderson_Maruf_2023}---to lobby against pro-driver regulations. In rare instances, platforms completely abandoned regions---e.g.,  Alaska (2014) \cite{Zak_2017}, San Antonio (2015) \cite{Griswold_2015}, Houston (2015) \cite{Mekelburg_2016}, Austin (2016) \cite{hampshire2017measuring}---until new (state) legislation passed which lifted platform regulations.

In an attempt to institutionalize their unregulated status across the U.S., platforms began and continue to lobby states over two main objectives. The first is pushing for state preemptive policies to override localized, city-initiated regulations which had begun attempting to restrict rideshare companies and protect drivers. The second is pressing states to declare drivers as independent contractors thereby concretizing their status as workers without benefits \cite{shererflexible}. This latter initiative allows platforms to reduce overhead cost and transfer risk and liability to workers. For example, drivers pay from their earnings for any fuel expended while working, whereas as employees, they would be reimbursed for expenses accrued while working. Additionally, drivers as independent contractors who are injured on the job will face monetary loss while unable to work, but as employees, they would be eligible for worker's compensation for a duration while unable to work.

Analysis from August 2019 by the Economic Policy Institute (EPI), a non-profit think tank that conducts research about the economic impact of policies, determined that at least 44 states preempt cities or counties in the regulation of rideshare companies and do not impose strict regulations themselves\footnote{large cities in certain states may be exempt such as New York City in New York and San Francisco in California}, and at least 34 states have passed legislation or ballot initiatives formalizing drivers as "nonemployees" or independent contractors \cite{shererflexible}. 

\subsection{Driver Organizing Efforts and Worker-Centered Regulation Attempts} \label{subsection_reports} Several driver organizations have formed to spur worker rights. Rideshare Drivers United, one of the largest with twenty thousand driver members, has been the driving force for California legislation. This includes the passage of AB5, which declared gig workers to be employees entitled to benefits, as well as the ongoing fight over Proposition 22, a platform-led ballot initiative which seeks to reverse AB5 \cite{dubal2020new, Browning_2023}. Other notable examples include New York Taxi Workers Alliance successfully pursuing minimum pay standard laws for rideshare drivers \cite{Ha_2018}; NYC's Los Deliveristas Unidos securing app delivery drivers a set of bills addressing minimum pay, transparent wages, and several well-being measures (e.g., the repurposing of NYC infrastructure as rest/recharge stations) \cite{krupat2023deliveristas}; Teamsters 117 pushing for ordinances around rideshare minimum pay and deactivation recourse in Seattle\footnote{These measures have since been replaced by Washington's statewide TNC regulatory requirements.} \cite{Zilly_2019}; Colorado Independent Drivers United advocating for state legislation that platforms provide fare transparency to drivers and passengers \cite{Cheshire_2023}; and the Chicago Gig Alliance organizing for safety standards, deactivation recourse, and driver wage policies \cite{Chicago_Gig_Alliance_2023} through the Rideshare Living Wage and Safety Ordinance.   

In some instances of driver organizing, policymakers have requested formal reports to support the need for and potential implications of related policy. An early example, in 2018, the New York City’s Taxi and Limousine Commission (TLC), tasked with regulating medallion cabs and for-hire vehicles including Uber and Lyft (commonly referred to as Transportation Network Companies or TNCs), was concerned that TNC drivers were earning less than minimum wage and commissioned researchers to investigate the need for and potential implications of proposed minimum pay standards \cite{parrott2018earnings}. Using administrative data provided to NYC by 4 major app-based companies, \citet{parrott2018earnings}'s findings showed drivers earning below minimum wage and hinted at how vulnerable the NYC rideshare driver population was---mostly full-time, immigrant drivers (60\%) dependent on the income for essential needs rather than part-time drivers working out of convenience for supplemental income as TNCs often state \cite{Stein_2020}. The TLC report importantly established 1) the standard wage metric as earnings \textit{post-expenses} and 2) that working time includes not just time spent transporting passengers, but the entirety of their work, e.g., commuting, driving to passengers, and circling blocks while resting or anticipating a work assignment \cite{parrott2018earnings}. These are important distinctions as the underlying assumptions about earnings, expenses, and time directly frame understanding of whether worker wages are livable.

Additional cities and states have proposed minimum pay policies or worker classification determinations, and similar economic analyses have been conducted to determine drivers' earnings under such policies and the impact on customers and the regional economy \cite{reich2020minimum, jacobs2021massachusetts, reich2020pay, Jacobs_Reich_2019, Manzo_Bruno_2021, henao2017impacts, zoepf2018economics, zipperer2022national, shierholzepi, schmitteconomic}. These reports have found that on average, drivers earn sub-minimum wages in the absence of policy. For example, \citet{reich2020minimum}, commissioned by the City of Seattle to inform a minimum pay ordinance, surveyed 30k workers and found drivers on average earn \$9.73/hr after expenses, with most rides completed by full-time drivers for whom the work is their primary source of income. The Mayor's Office conducted its own extensive driver outreach through townhalls, interviews, focus groups, and surveys of over 10k drivers, finding the majority of respondents were non-white, two-thirds drive for a TNC as their only job, and drivers face extreme pressures due to rising expenses, limited flexibility, and uncertainty inherent to platform work \cite{Seattle_2020}. Platform-based policies to classify workers as independent contractors have also been scrutinized such as the EPI comparing independent contractors with in-house employee counterparts, revealing the independent contractor classification of gig workers risks underpayment \cite{zipperer2022national, shierholzepi, schmitteconomic}. Advocacy and driver-led organizations have released their own reports, similar to \cite{reich2020comparison, parrott2018earnings}, to provide more context into local conditions and worker experiences \cite{Leverage_Dalal_2022, washington2019delivering, McCullough_Dolber_2021, McCullough_2022, figueroa2021essential, steward2023gig}. This includes \citet{McCullough_2022}'s driver-led initiative to collect and analyze driver data finding median driver wages to be \$6.20/hr under Prop 22 and \citet{Leverage_Dalal_2022}'s findings of similar sub-minimal wages in Denver and understanding about the driver population being primarily full-time and people of color.

Unsurprisingly, platforms have pushed back on policy and regulation attempts \cite{hyman2020platform, thornberg2020proposition}. On the same day \citet{reich2020minimum}'s Seattle report was published, an Uber-Lyft sponsored study by \citet{hyman2020platform} was released, claiming median driver earnings at \$23.25/hr. Researchers have countered these \cite{jacobs2020effects, reich2020comparison, mishel2018uber}, such as \citet{reich2020comparison} breaking down the assumptions used by \citet{hyman2020platform} that lead to inflating driver earnings, such as 1) including tips in their calculations and disregarding time drivers spend waiting to receive an assignment, and 2) underestimating driver expenses by assuming most are casual, part-time drivers when \cite{Seattle_2020} and \cite{reich2020minimum} both found contradictory patterns in Seattle. Though surveys have shown that nationwide most gig workers\footnote{In these surveys, gig workers include responses from app-based workers and others who are independent contractors as opposed to in-house employees.} work part-time \cite{garin2023evolution, anderson2021state}, \cite{anderson2021state} found that gig work income remains essential or important to meeting basic needs for most workers, and separately, \cite{zipperer2022national} found 29\% of respondents made below their state's minimum wage.



\section{Charting Future Directions for HCI and Policymaking Collaborations} \label{Future}
We briefly expand on two ideas as they relate to supporting HCI and policy collaborations: 1) data advocacy and transparency initiatives and 2) legal/regulatory approaches. Note: this section is not intended to be comprehensive of all potential future directions for HCI and policymaking. Instead, it is meant to be \textit{informative} as a starting point to direct researchers to helpful resources they could build off of, and \textit{speculative} to encourage scholars to consider different provocative questions or angles which emphasize interdisciplinary collaborations or perspectives.

\subsection{Data Advocacy \& Transparency Initiatives}
A growing interest amongst HCI and CSCW researchers is how to use data to \textit{investigate} and \textit{audit} technology companies. In lieu of standards or policies for responsible technology development and practices, \textit{algorithm auditing} \cite{sandvig2014auditing} originated as a (research) method to investigate tech companies' "black-box" algorithms. This involves the "collection and analysis of outcomes from a fixed algorithm or defined model within a system...[often to]...uncover problematic patterns in models of interest" \cite{raji2019actionable}. Algorithm auditing has subsequently been used to reveal how algorithms are exacerbating discrimination, distorting information to different users, exploiting users' sensitive information, or making incorrect predictions or classifications \cite{bandy2021problematic}. Prominent cases of algorithm auditing include \citet{buolamwini2018gender}'s work finding that commercial datasets for facial recognition objectively perform worse on darker-skinned females, signifying gender and racial disparities in classification accuracy of technologies; and \citet{angwin2016machine}'s reporting of how a US county's recidivism prediction algorithm led to discriminatory racial disparities in bail determinations. Other domains that researchers have explored with algorithm auditing and uncovered race or gender discrimination include housing loan determinations \cite{martinez2021secret, asplund2020auditing, bhutta2022much}, child welfare risk predictions \cite{cheng2022child}, search engine results \cite{robertson2018auditing, noble2013google, makhortykh2021hey}, and social media recommendation algorithms \cite{srba2023auditing}.

Much of the research examining worker experiences under algorithmic management highlights qualitative stories, e.g., elevating drivers' accounts with algorithmic management through rich narratives or discussions that drivers have shared about working for rideshare platforms. While this information format is indispensable, research on civic engagement in local policymaking processes demonstrates that policymakers consistently value data over strict anecdotes \cite{manuel2020place, spaa2019understanding, manuel2017participatory}. Under these circumstances and given the recent interest from the federal government to investigate things like platform wage practices \cite{FTC_2022}, it may be beneficial to turn to research on \textbf{algorithm auditing}---particularly \textbf{user-driven algorithm auditing}---for how it can support policymaking that holds tech platforms accountable. Algorithm auditing is especially appropriate for tech policymaking because in the absence of wide reaching regulation to oversee potential harms of technology, researchers, activists, and even everyday users have engaged in this process to surface ongoing or potential harms that technologies enable.

Rideshare driver minimum pay reports described in Section \ref{subsection_reports} can be viewed as variations of algorithm auditing as these present analyses of drivers' actual pay under rideshare platform conditions \cite{parrott2018earnings, reich2020minimum, jacobs2021massachusetts}. However, these reports have been limited to establishing the need for minimum pay standard laws for rideshare drivers as opposed to protections against algorithmic management itself. And while early versions of these reports were able to utilize data provided by platforms themselves \cite{parrott2018earnings}, rideshare companies have since resisted providing this data for independent researchers to create similar reports about different cities \cite{Jacobs_Reich_2019, reich2020minimum}. 

While policy and regulation of gig work app platforms themselves lag at the national level in the US, the evolution of data privacy laws within several states (e.g., California Consumer Protection Act; New York City's Privacy Act, Freedom of Information Law, Open Data Law) and globally (e.g. European Union's General Data Protection Regulation; Brazil's  Lei Geral de Proteção de Dados, LGPD) offer paths forward for supporting \textbf{worker-led algorithm auditing}, a form of user-driven algorithm auditing. Under these laws, platforms are required to make available for workers access to downloading their individual work data collected by platforms. Some platforms have also made some or all of this work data accessible for download by workers in regions not currently covered by data privacy laws. This along with research that seeks to empower people using their data as tools of resistance against tech platforms \cite{vincent2021data, li2023dimensions, li2018out, hsieh2024data}, as well as organizations such as Driver's Seat Cooperative (US)\footnote{https://driversseat.co/, now merged with Workers' Algorithm Observatory (US) https://wao.cs.princeton.edu/} and Worker Info Exchange(UK)\footnote{https://www.workerinfoexchange.org/} that help workers in collecting and analyzing work data strengthens the possibilities of worker-led algorithm auditing. 

This suggests that one promising avenue for HCI researchers to pursue is how to support the collection, analysis, and presentation of worker data towards initiatives for gig worker well-being. This may consist of creating and testing tools that enable the recruitment and collection of worker data for specific auditing purposes---e.g., \citet{calacci2022bargaining}'s collaboration with a non-profit worker collective to audit a gig work platform's pay algorithm. This can include exploring creative formats for worker data---such as data probes \cite{zhang2023stakeholder}---which enable workers to explain how their data patterns reflect platform tactics to control workers. \citet{zhang2024data} also found that these data probes can illuminate worker conditions and platform practices for non-rideshare drivers, such as policy-adjacent stakeholders. And lastly, this can look like tools that closely link workers and policymakers---for example, \citet{hsieh2024data} have proposed "worker data collectives" where the goal is to not only help workers pool their data for investigation, but to make this anonymized, aggregated information available for policymakers and regulators to view so they understand issues workers face that require lawmaking or regulatory action.


\subsection{Exploring Legal Frameworks}
During our research for \cite{zhang2024data}, we came across relevant frameworks and legal processes that we were previously unaware of or had not drawn connections to. This led us to speculate how HCI/CSCW researchers can complement such processes, or how certain proposed frameworks can help bridge HCI and policy research efforts. 

Legal scholar \citet{dubal2023algorithmic} offers a helpful lens for understanding worker experiences and holding platforms accountable for labor practices. She explains that platforms' data extraction of workers' labor allows them to create opaque, algorithmically determined and personalized pay and incentives, engendering \textbf{algorithmic wage discrimination}, a troubling practice that violates worker autonomy, reduces wages, and risks reinscribing social and racial equity issues. For example, based on a worker's history of ride acceptances, platforms can offer personalized pricing that caters to the lowest a worker is willing to accept for a task. Thus, it is possible to offer different wages for the same work (here, a ride request) to Worker A vs. Worker B. Without policy or regulation to limit this practice and lacking data disclosure policies to shed light on when/how this variable pay is occurring, platforms can implement this for workers and even riders. Similar to the driver example above, Rider A and Rider B can hypothetically be offered different ride prices for the same or similar trip based on differences in their historical trip price acceptance threshold. Dubal frames this as violating fair employment models and argues for a ban on these practices. To that end, we encourage collaborations between HCI and Law which directly investigate this phenomenon---especially with an angle on whether platforms are violating existing U.S. anti-discrimination protections such as Equal Pay through "disparate impact". A number of universities have established Technology, Law, \& Policy Clinics (e.g., University of Washington\footnote{https://www.law.uw.edu/academics/experiential-learning/clinics/technology-law}, KU Leuven\footnote{https://www.law.kuleuven.be/legal-clinic-ai-and-human-rights}) and this may be a promising case study for interdisciplinary teams of these clinics to tackle. 

Additionally, we draw a connection between algorithmic wage discrimination as experienced by workers and algorithmic pricing as experienced by riders. In May 2024, the FTC's Office of Technology published a blog post delineating 8 research questions they are currently exploring\footnote{https://www.ftc.gov/policy/advocacy-research/tech-at-ftc/2024/05/p-np-not-exactly-here-are-some-research-questions-office-technology}. One of these is "algorithmic pricing"---otherwise defined in the post as "price fixing by algorithms". This correlates with the variable pay that gig workers describe experiencing. Collaboration here between the FTC Office of Technology and academics/practitioners creating tools or conducting research to surface evidence of algorithmic wage discrimination or algorithmic pricing---including worker-driven auditing research---holds promise not just to protect gig workers but also the consumers of the work (i.e., those requesting services). 

Recent scholarly research, federal rulemaking, and advancements in state legislation for gig work data transparency suggest complementary paths to visit. First, in lieu of the U.S. mandating data disclosure from digital platforms for workers or consumers, \citet{rao2024rideshare} calls for platforms to provide regular releases of transparency reports which offer insights into metrics such as "individual ride statistics, driver history, and details about platform algorithms' inputs and outputs". Second, in early 2024, the Department of Labor revised their guidance on how to classify workers as independent contractors or employees\footnote{https://www.dol.gov/agencies/whd/flsa/misclassification/rulemaking} to more closely scrutinize the degree to which workers can practice economic independence from "employers" \cite{Lubart_2024}. A supportive research undertaking might be analyzing worker data against these classification rules computationally to identify instances of violations to the independent contractor status. Finally, at the state level, the Colorado legislature introduced legislation to provide more protections and transparency for rideshare drivers \cite{Colorado_General_Assembly_2023} which the governor signed in June 2024 \cite{Colorado_General_Assembly_2024}. This is one of the first of its kind, with language that notably avoids discussing the (often unresolved) topics rideshare policy or lawsuits often focus on---worker classification and minimum pay standards. Once implemented, it may be possible to leverage new data disclosures towards investigations detailed above such as algorithmic wage discrimination and violations of worker classification.

\section{Conclusion}
In this paper, we explored the following questions:

\begin{enumerate}
    \item What challenges exist for policymakers and regulators when overseeing rideshare platforms?
    \item What HCI opportunities exist for advancing policymaking of app-based rideshare companies?
\end{enumerate}

To answer the first, in Section \ref{Section_Background_Rideshare} we explained one challenge is how platforms often create specialized terminology for their features to promote the idea that they differ from their regulated counterpart, the taxicab industry. In Section \ref{Section_algorithmic_management}, we explained how workers have actually experienced these features as heavy-handed management and control, a contradiction to their independent contractor status. In Section \ref{Section_policymaking}, we showed that another challenge is the opposing policy interests of platforms and workers by contrasting platform strategies to solidify their unregulated status with the policy goals workers have attempted to push forward.

To answer the second question, we described two directions of future work for researchers seeking to combine HCI and policymaking efforts: (1) supporting data transparency initiatives and (2) more closely aligning HCI contributions with legal, policy, and regulatory frameworks. Around the former, we expanded on worker-led algorithm auditing. For the latter, we commented on various legal and regulatory frameworks for HCI researchers to consider. Though in our paper, we focused on one use case---rideshare platforms---we believe that several ideas presented in this paper do not have to be limited to rideshare-related tech policymaking. For example, principles for worker-led auditing can be applicable to other digital platform gig workers (e.g., couriers, food delivery workers, petsitters). Additionally, given the increasing data collection on workers in traditional workplaces \cite{ajunwa2017limitless, ball2022surveillance, bowen2017personal, kresge2020data}, the considerations we have presented here about how platforms use worker data to coerce behaviors or impose punishments will be relevant when designing technology policy more broadly. 

Lastly, we share additional resources we have compiled to support the work of other researchers and advocates in this space.

\begin{enumerate}
    \item First, a companion website to this paper. This informational website is intended for any reader---general public, researcher, policymaker, journalist, etc. We invite you to explore and reach out to us for comments and questions. 
    
    \href{https://www.demystifying-gigwork.com}{\color{blue}Website: demystifying-gigwork.com}.
    
    \item Second, a compilation of related rideshare policy/regulation reports by worker-organizer groups, economists, governments, and platforms. This space is fast moving and this database has been manually compiled. However, these are helpful resources that we found and wanted to share with anyone interested in learning more. 
    
    \href{https://airtable.com/app09qGH7CDmoN6qE/shrQEYxZjSUYQrZy7/tblUBrABNUUToPVfN}{\color{blue}Link to Repository of Reports}.
    \item Third, a timeline of rideshare (and relevant app-based gig work) policy/regulation related events in the U.S. While this is abbreviated and is only current through late 2023, it may be helpful for someone just getting started. We personally found the space to be somewhat tangled and disparate, and needed a way to keep track of local and state efforts and outcomes.
    
    \href{https://airtable.com/appZvP15Nt6EAjV8R/shrBZ9AKDhJ7woa9s/tblVguO09iLEE18o8}{\color{blue}Link to Timeline}.
    \item Lastly, a glossary of helpful gig work and rideshare specific terms. Unsurprisingly, there are many specialized terms used when describing and understanding gig work, rideshare platforms, and policy. We found this to be the case when rideshare participants and policy-related stakeholders had to clarify terms they were using with us; and when we ourselves were seeking how to best explain this work to researchers not already embedded in the gig work or rideshare domain. 

   \href{https://airtable.com/appMlzch3KfhPzK2R/shrKrHSstyAIS3usV/tblQ1XsbYafev121J}{\color{blue}Link to Glossary}.
\end{enumerate}

\begin{acks}
Heartfelt thanks to my advisor, Min Kyung Lee, and co-authors across the course of this work, including Alexander Boltz, Rocita Rana, Chun-Wei Wang, Jonathan Lynn, and Veena Dubal. I'm also very grateful to the students I have had the good fortune to work with through the Human-AI Interaction Lab led by Dr. Min Kyung Lee, including those who have worked on the companion website---Meah Lin, Jake Lohman, Nithila Sathiya, and Praveen Mogan. The research informing this paper was partially supported by the following: the National Science Foundation CNS-1952085, IIS-1939606, DGE-2125858
grants; Good Systems, a UT Austin Grand Challenge for developing responsible AI technologies; and UT Austin’s School of
Information. 
\end{acks}

\bibliographystyle{ACM-Reference-Format}
\bibliography{references}



\end{document}